\documentclass[aps,prb,twocolumn,groupedaddress,showpacs]{revtex4}

\usepackage{epsfig}

\begin{document}


\title{Normal-Superconducting Phase Transition Mimicked by Current Noise}

\author{M.~C. Sullivan}
\author{T. Frederiksen}
\author{J. M. Repaci}
\author{D. R. Strachan}
\author{R. A. Ott}
\author{C. J. Lobb}
\affiliation{Center for Superconductivity Research, Department of
Physics, University of Maryland, College Park, MD 20742}



\begin{abstract}
As a superconductor goes from the normal state into the
superconducting state, the voltage vs. current characteristics at
low currents change from linear to non-linear. We show
theoretically and experimentally that the addition of current
noise to non-linear voltage vs. current curves will create ohmic
behavior. Ohmic response at low currents for temperatures below
the critical temperature $T_c$ mimics the phase transition and
leads to incorrect values for $T_c$ and the critical exponents
$\nu$ and $z$. The ohmic response occurs at low currents, when the
applied current $I_0$ is smaller than the width of the probability
distribution $\sigma_I$, and will occur in both the zero-field
transition and the vortex-glass transition. Our results indicate
that the transition temperature and critical exponents extracted
from the conventional scaling analysis are inaccurate if current
noise is not filtered out.  This is a possible explanation for the
wide range of critical exponents found in the literature.
\end{abstract}


\maketitle


The occurrence of a wide critical regime of the high-temperature
superconductors\cite{chris} -- and the subsequent theories
regarding the phase transition that occurs in this
regime\cite{ffh} -- have led many researchers to look for critical
behavior in the non-linear voltage vs. current ($I-V$)
characteristics of these superconductors.\cite{see-doug}  This
behavior has been studied in many materials in a variety of
different conditions.  The most widely researched material is
$\mathrm{YBa_{2}Cu_{3}O_{7-\delta}}$ (YBCO): thick films
(thickness $d \approx 2500$~\AA),\cite{consensus0} thin films ($d
< 1000$~\AA),\cite{thinfilms} and bulk single crystals.\cite{yeh1}
YBCO has been measured both in a magnetic field (the vortex-glass
or Bose glass transition) and in zero field.\cite{lowfields}
Researchers have also investigated the vortex-glass transition in
(to name but a few):
$\mathrm{Bi_{2}Sr_{2}CaCu_{2}O_{8+\delta}}$,\cite{bscco}
$\mathrm{Nd_{1.85}Ce_{0.15}CuO_{4-\delta}}$,\cite{ncco} other more
unusual superconductors,\cite{exotic} and critical behavior has
even been reported in some low-$T_c$ systems.\cite{lowtc}  This
large body of work has led to the general consensus that the
vortex-glass transition exists, despite some arguments to the
contrary.\cite{against} However, there is a wide range of reported
critical exponents $\nu$ and $z$ from the experimental $I-V$
curves. Our recent work has called into question the validity of
the conventional scaling analysis,\cite{doug} as we demonstrated
multiple data collapses, each with its own set of critical
parameters, using only one set of experimental data.

In this report we discuss the under-appreciated and invidious
behavior of current noise when measuring non-linear $I-V$ curves.
The normal-superconducting phase transition manifests itself at
low currents as a change from ohmic behavior ($T > T_c$) to
non-linear behavior ($T < T_c$).  We show, both theoretically and
experimentally, that the addition of current noise to a device
with an intrinsic non-linear response will create an ohmic
response at low currents.  Thus, current noise will create ohmic
behavior at low currents even for temperatures below $T_c$, and
isotherms that are actually \textit{below} $T_c$ will appear to be
\textit{above} $T_c$. In this manner, current noise will mimic the
phase transition, and will lead to an underestimate of $T_c$, and
incorrect values for $\nu$ and $z$ -- and in the worst case, the
ohmic response due to noise will give the impression that the
phase transition does not exist. This will occur both in zero and
non-zero field. Thus, different amounts of current noise (highly
dependant on the experimental setup) will lead to different values
for the critical exponents (expected to be universal).  This
effect, especially when combined with the flexibility inherent in
scaling,\cite{doug} is a possible explanation for the many
different critical exponents reported in the literature.

To understand this effect more fully, we look at the underlying
equations.  When measuring the $I-V$ curves of superconductors, we
apply a dc current $I_0$ and measure the average voltage, $\langle
V \rangle$. Let us suppose that at some temperature $T$ the sample
has a response $V = f(I)$, where $f(I)$ can be non-linear in
current, and $f(I) = -f(-I)$, i.e., anti-symmetric. Because any
applied current will have noise (which may be shot noise, Johnson
noise, 1/f noise, or noise from external sources such as the
electronics), the measured voltage $\langle V \rangle$ will be
given by\cite{theses}
\begin{equation}
\langle V \rangle = \int_{-\infty}^{\infty}
f(I)P(I-I_0)dI,\label{eq:v-meas}
\end{equation}
where $P(I-I_0)$ is the probability distribution for the current,
which is centered about the applied current $I_0$.  We assume
$P(I-I_0)$ is symmetric about $I_0$, as there is no preferred
direction for current noise to flow. $P(I-I_0)$ has a width
$\sigma_I$ given by the variance of the probability distribution,
$\sigma_I^2 = \int_{-\infty}^{\infty} (I-I_0)^2 P(I-I_0) dI$.

When $I_0 \gg \sigma_I$, the distribution is very narrow, and only
values of $f(I)$ within a few $\sigma_I$ will contribute to
$\langle V \rangle$ in Eq.\ \ref{eq:v-meas}.  We expand $f(I)$ in
a Taylor series to find $f(I) = f(I_0) + f'(I_0)(I-I_0) +
\frac{1}{2}f''(I_0)(I-I_0)^2 + \cdots$.  When inserted back into
Eq.\ \ref{eq:v-meas}, due to symmetry, only the even terms in the
expansion contribute, thus\cite{theses}
\begin{equation}
\langle V \rangle = f(I_0) + \frac{1}{2} f''(I_0) \sigma_I ^2 +
\cdots,
\end{equation}
and we see that
\begin{equation}
\langle V \rangle \approx f(I_0), \ \mathrm{for} \ I_0 \gg
\sigma_I.\label{eq:high-i}
\end{equation}
Thus, the finite width of $P(I-I_0)$ has no effect when the
applied current $I_0$ is much larger than the noise current
$\sigma_I$, and the measured voltage is independent of the noise.

The situation is markedly different when $I_0 \ll \sigma_I$. In
this case, because $I_0$ is small, we will expand the probability
distribution about $I$: $P(I-I_0) = P(I) - I_0 \frac{\partial
P(I-I_0)}{\partial (I-I_0)}|_{I_0 = 0} + \mathcal{O}(I_0^2)$. When
this distribution is inserted back into Eq.\ \ref{eq:v-meas}, we
find that the first term does not contribute due to symmetry and
thus, to first order in $I_0$,\cite{theses}
\begin{equation}
\langle V \rangle \approx I_0 R_{eff}, \ \mathrm{for} \ I_0 \ll
\sigma_I \label{eq:ohmic_tail}
\end{equation}
where $R_{eff}$ is an effective resistance given by
\begin{equation}
R_{eff} = - \int_{-\infty}^{\infty}f(I)\frac{\partial
P(I)}{\partial I}dI.\label{eq:ohmic-tail}
\end{equation}
This means that, if $I_0 \ll \sigma_I$, the measured voltage is
\textit{always} linear in the applied current, independent of the
form of $f(I)$! Even strongly non-linear $I-V$ curves will appear
ohmic at low currents.\cite{doug-note}  This occurs both above and
below $T_c$, and will occur in zero field as well as in the
vortex-glass transition.

This ohmic response at low currents is especially damaging because
it can mimic the true ``ohmic tails" expected for $T<T_c$ in a
phase transition. For $T>T_c$, as $I \rightarrow 0$ it is
predicted that (for D=3)\cite{ffh}
\begin{equation}
\frac{V}{I} \sim \left(\frac{T-T_c}{T_c}\right)^{\nu(z-1)},
\label{eq:rl}
\end{equation}
where $\nu$ is a static critical exponent and $z$ is the dynamic
critical exponent. Thus, an ohmic tail generated by noise via Eq.\
\ref{eq:ohmic_tail} can be easily mistaken for the ohmic tail
expected from the phase transition in Eq.\ \ref{eq:rl}, especially
as they are both predicted to happen at low currents.

In general, $\langle V \rangle$ and $R_{eff}$ are impossible to
determine analytically because the function $f(I)$ is unknown. The
form of $f(I)$ is known in two regions: the normal state, and at
$T_c$.  In the normal state, the sample is a simple resistor, such
that $V = I R_0$.  At $T_c$, the voltage is expected to be a power
law in current, such that $V = b I^a$,
where the exponent $a$ incorporates the dynamic exponent $z$
($a=\frac{z+1}{2}$ for D=3).\cite{ffh}

We can determine $\langle V \rangle$ for these two cases.  We
assume a Gaussian form for $P(I)$, since we expect the noise
fluctuations in the leads to be the result of the (almost) random
motion of a huge number of electrons (stochastic motion), such
that
\begin{equation}
P(I-I_0) = \frac{1}{\sigma_I \sqrt{2 \pi}} e^{-(I-I_0)^2/2
\sigma_I^2}.\label{eq:prob-i-gauss}
\end{equation}
We can then insert $f(I)$ and this form for $P(I-I_0)$ into Eq.\
\ref{eq:v-meas} to find $\langle V \rangle$. When $V=f(I) = I R_0$
(in the normal state, or at low currents when $T>T_c$ in the
critical regime), we find
\begin{equation}
\langle V \rangle = I_0R_0, \ \ \mathrm{in\ the\ normal\
state}\label{eq:v-meas-norm}
\end{equation}
as expected for a simple resistor.\cite{doug-note2}  On the other
hand, at $T_c$ when $V=f(I)=bI^a$, we find at low currents that
the measured voltage is linear in the applied current, $\langle V
\rangle = I_0 R_{eff}$, where $R_{eff}$ is given by
\begin{equation}
R_{eff} = b \sigma_I^{a-1} \sqrt{\frac{2^{a+1}}{\pi}} \cdot
\Gamma(\frac{a}{2} +1),\ \mathrm{for} \ T=T_c,  \label{eq:R-eff2}
\end{equation}
and $\Gamma$ is the gamma function.

For a given experimental $I-V$ curve which is non-linear at high
currents and ohmic at low currents, we can fit its high-current
behavior to a power law to find $a$ and $b$, and its low-current
ohmic tail to find $R_{eff}$.  If we assume the ohmic tail is
entirely caused by noise, we can estimate the noise necessary to
create the ohmic tail, as
\begin{equation}
\sigma_I = \left[\frac{R_{eff} \sqrt{\frac{\pi}{2^{a+1}}}}{b
\Gamma(\frac{a}{2} +1)}\right]^{\frac{1}{a-1}}.\label{eq:sigmaI}
\end{equation}
We can compare this estimate with the noise as measured with a
spectrum analyzer.

We have examined the phase transition in zero field using current
vs. voltage ($I-V$) curves of $\mathrm{YBa_{2}Cu_{3}O_{7-\delta}}$
(YBCO) films deposited via pulsed laser deposition onto
$\mathrm{SrTiO_3}$ (100) substrates.
 X-ray diffraction verified that our films are of predominately
c-axis orientation, and ac susceptibility measurements showed
transition widths $\leq 0.25$ K. $R(T)$ measurements show $T_c
\approx 91.5$ K and transition widths of about 0.7 K.  AFM and SEM
images show featureless surfaces with a roughness of $\approx 12$
nm. These films are of similar or better quality than most YBCO
films reported in the literature.

In preparation for measurement, we photolithographically patterned
our films into 4-probe bridges of width 8 $\mathrm{\mu}$m and
length 40 $\mathrm{\mu}$m and etched them with a dilute solution
of phosphoric acid without noticeable degradation of $R(T)$.  We
surround our cryostat with $\mu$-metal shields to reduce the
ambient field to $2 \times 10^{-7}$ T, as measured with a
calibrated Hall sensor.  We routinely achieve temperature
stability of better than 1 mK at 90 K. To reduce noise, our
cryostat is placed inside a screened room and all connections to
the apparatus are made using shielded tri-axial cables.

We have experimented with several different filtering schemes.  We
use only passive filters, so as not to introduce noise from an
active filter. Our typical filtering scheme, similar to others
reported in the literature,\cite{yeh1} uses low-pass $\pi$ filters
(insertion loss of 3 dB at 4 kHz) at the screen room wall. We have
also used low-pass T filters (3 dB at 2 kHz) and double-T filters
(3 dB at 2 kHz with a sharper cutoff) at the top of the probe.
Additionally, we modified our probe to accept filters at the cold
end.  At 90 K, and the 3 dB point of the low-pass T filters shifts
upwards to 70 kHz.  We also used cold copper-powder
filters\cite{cu-powder} that have a measured insertion loss
greater than 60 dB for frequencies greater than 5 GHz.

\begin{figure}
\centerline{\epsfig{file=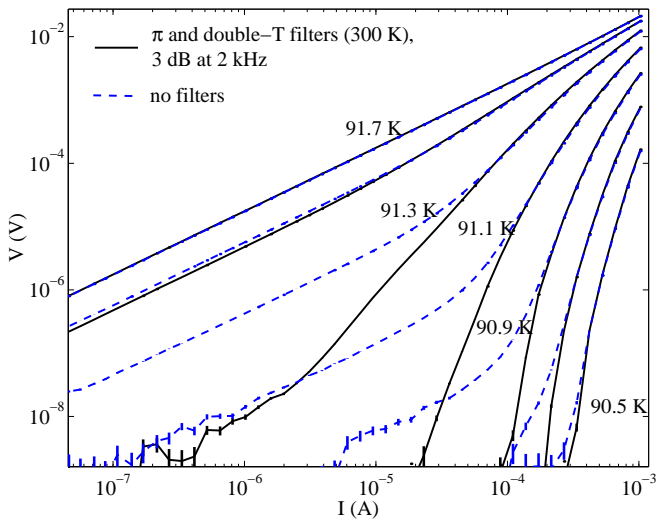,clip=,width=\linewidth}}
\caption{Two sets of $I-V$ curves for a 2100~\AA~thick film with
bridge dimensions $8\times40\ \mu$m$^2$. The solid lines are
isotherms taken with low-pass $\pi$ filters (3 dB at 4 kHz) at the
screened room wall and low-pass double-T filters (3 dB at 2 kHz)
at the top of the probe. The dashed lines are isotherms taken
without filtering. The isotherms are separated by 200 mK, and the
error bars are shown (when larger than the lines). The
highest-temperature isotherms (91.7 K) are ohmic, and fully in the
normal state.  In the transition region, we see (especially in the
isotherms at 91.1 K and 90.9 K) that noise creates ohmic tails in
non-linear signals.} \label{fig:noise1}
\end{figure}

The theoretical prediction that noise creates ohmic tails is
easily seen experimentally.  We can dramatically increase the
amount of noise in our system by removing the filters and leaving
the door to our screened room open.  We can then compare isotherms
with and without filtering.  Two sets of $I-V$ curves for one
sample are shown in Fig.\ \ref{fig:noise1}.

In this figure, the highest-temperature isotherm (91.7 K) is in
the normal state, and has a slope of one (indicating ohmic
behavior, $V \sim I$). In the transition region, the effect of
noise is dramatically apparent.  The filtered isotherms (solid
lines) and the unfiltered isotherms (dashed lines) overlap at high
currents, as predicted by Eq.\ \ref{eq:high-i}, indicating that
the additional noise has no effect.  At lower currents, however,
the unfiltered isotherms deviate and become ohmic (same slope as
the isotherm at 91.7 K), as expected from Eq.\
\ref{eq:ohmic-tail}. This effect is most noticeable in the
isotherms at 91.1 K and 90.9 K, where the non-linear isotherms
become ohmic at low currents when the filters are removed.

It is also easy to see how these ohmic tails due to noise could be
mistaken for ohmic tails due to the 3-D phase transition.  The
ohmic tail due to noise at 90.5 K drops below the resolution of
our voltmeter (1 nV), thus the unfiltered isotherm appears
non-linear. This transition from isotherms with an ohmic tail
(91.1 K and 90.9 K) to (apparently) non-linear isotherms (90.5 K
and below) is the same signature we expect from the phase
transition.  From the unfiltered isotherms alone, the conventional
analysis of $I-V$ curves would lead us to say that $T_c \approx
90.5$ K, despite the fact that the ohmic tails at 91.1 K and 90.9
K are artifacts created by noise.  Note also that the filtered and
unfiltered isotherms are equally smooth.  Once the noise reaches
the sample, its response changes, thus the measured isotherm will
appear smooth, regardless of how much noise is in the system.

In an attempt to further filter our leads, we added T filters and
copper-powder filters\cite{cu-powder} to the cold end of our
probe, very close to the sample. Isotherms taken with warm
filtering at the screened room wall and the top of the probe
(solid lines in Fig.\ \ref{fig:noise1}) were identical to
isotherms taken with filters at the screened room wall, top of the
probe, and at the cold end of the probe.  Thus, the addition of
cold filters did not improve the data.  From this we conclude that
the Johnson noise created in the probe wiring is not significant.

\begin{figure}
\centerline{\epsfig{file=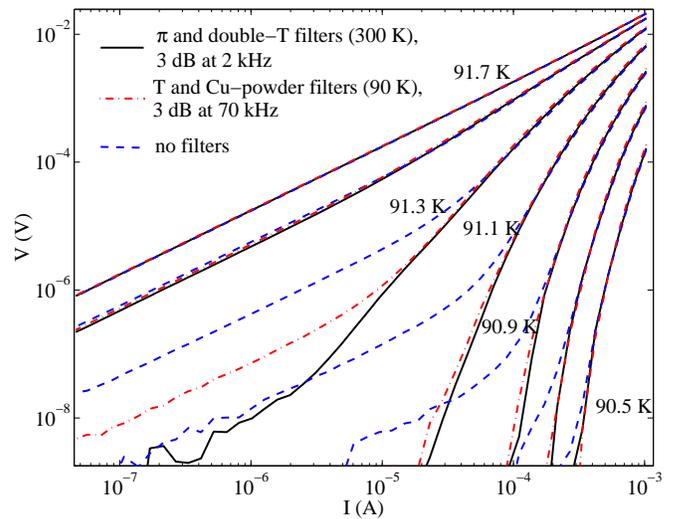,clip=,width=\linewidth}}
\caption{Three sets of $I-V$ curves for one sample.  The solid
lines and the dashed lines are the same isotherms from Fig.\
\ref{fig:noise1}.  The dotted-dashed lines are isotherms taken
with T filters and copper-powder filters at the cold end of the
probe (3 dB at 70 kHz). The isotherms are separated by 200 mK, and
in this figure, the error bars are suppressed for clarity. We can
see that a 3 dB point of 70 kHz is not low enough to filter the
isotherm completely.} \label{fig:noise2}
\end{figure}

It is instructive to consider the effect of cold filters alone vs.
warm filters alone.  Because the 3 dB point of the T filters
shifts to 70 kHz when cold, we can compare low-pass filters with
different 3 dB points.  Isotherms taken with all three filter
configurations are shown in Fig.\ \ref{fig:noise2}.  From Fig.\
\ref{fig:noise2} it is obvious that a 3 dB point of 70 kHz is not
low enough to filter the noise properly, although the difference
between warm filters and cold filters is only obvious at 91.3 K.

This result leads us to wonder whether even a 3 dB point of 2 kHz
is low enough to properly filter the data.  Commercial passive
filters with a 3 dB point lower than 2 kHz are hard to find, but
we can resolve this question using another method.  The
environment connected to the sample generates a certain amount of
current noise, but the $I-V$ curves depend not on current but
rather current density. Therefore, if we test 4-probe bridges of
different widths, for a given amount of noise \textit{current}, we
can reduce the noise \textit{current density} using wider bridges.
We expect bridges of different widths to have similar $J-E$
curves, where $J$ is current density and $E$ electric field.
However, if noise is still a problem, wider bridges should show
different $E-J$ curves. We have measured bridges of different
widths,\cite{me} and have found that for typical filtering (3 dB
at 2 kHz), the $E-J$ curves for bridges of different widths (and
thus different noise current densities) are identical, indicating
that our low-pass $\pi$ filters are sufficient filtering.

Additionally, we can take an isotherm and use Eq.\ \ref{eq:sigmaI}
to estimate the amount of current noise required to create the
ohmic tail. If we take the filtered isotherm at 91.3 K from Fig.\
\ref{fig:noise1}, we find from the high currents $a \approx 1.55$
and $b \approx 10^{2.8}$ V/A$^a$. From the low currents, we find
$R_{eff} \approx 0.7\ \Omega$.  We can plug this in to Eq.\
\ref{eq:sigmaI}, and find $\sigma_I \approx 1.3\ \mu$A, if the
ohmic tail were caused by noise. We have measured the noise in our
probe using a spectrum analyzer and found $\sigma_I \leq 10$ nA,
far less than the estimate from Eq.\ \ref{eq:sigmaI}, indicating
that, with proper filtering, noise does not create the ohmic
tails.

Finally, it is interesting to note that we can change the
resistance of the ohmic tail at 91.3 K by adding noise in Fig.\
\ref{fig:noise2}.  We know from Eq.\ \ref{eq:v-meas-norm} that
adding noise to a linear $I-V$ curve does \textbf{not} change the
resistance.  This result indicates that the underlying behavior at
low currents of the 91.3 K isotherm \textit{must} be non-linear!
The ohmic tail that occurs even in the filtered data must result
from some other effect.  In Ref.\ \onlinecite{me}, we argue that
this occurs due to the finite thickness of our films.

We have shown, theoretically and experimentally, that the addition
of current noise can create ohmic behavior at low currents in
non-linear $I-V$ curves.  We have also shown that, in our
experimental setup, passive low-pass $\pi$ filters eliminated the
effects of noise. However, without filters, it is easy to confuse
ohmic tails generated by noise with ohmic tails expected from the
phase transition, causing incorrect choices of $T_c$, $\nu$, and
$z$.  These exponents are expected to be universal, though many
different exponents are reported in the literature. Filtering
schemes are rarely explicitly mentioned in the literature, and
thus current noise may be a possible explanation for the lack of
consensus regarding the exponents.

The authors thank J.~S. Higgins, D. Tobias, A.~J. Berkley, S. Li,
H. Xu, M. Lilly, Y. Dagan, H. Balci, M.~M. Qazilbash, F. C.
Wellstood, and R.~L. Greene for their help and discussions on this
work. We acknowledge the support of the National Science
Foundation through Grant No. DMR-0302596.




\begin{thebibliography}{99}
\bibitem{doug-note}
Eq. \ref{eq:ohmic_tail} is always true for any symmetric
$P(I-I_0)$ and antisymmetric $f(I)$, given that $I_0 \ll
\sigma_I$.
\bibitem{doug-note2}
Again, Eq. \ref{eq:v-meas-norm} is true for any symmetric
$P(I-I_0)$.
\end{thebibliography}
\end{document}